\documentclass[a4paper]{article}
\usepackage{graphicx}
\usepackage{amsmath}
\usepackage{hyperref}
\usepackage{color}

\begin{document}

\begin{center}

{\bf \Large Scaling of spin avalanches in growing networks }\\[5mm]

{\large Joanna Tomkowicz$^*$ and Krzysztof Ku{\l}akowski$^+$ }\\[3mm]

{\em

Faculty of Physics and Applied Computer Science,

AGH University of Science and Technology,

al. Mickiewicza 30, PL-30059 Krak\'ow, Euroland

}


{\tt ($^*$tomkowicz,$^+$kulakowski)@novell.ftj.agh.edu.pl}

\bigskip

\today

\end{center}

\begin{abstract}
Growing networks decorated with antiferromagnetically coupled spins are archetypal
examples of complex systems due to the frustration and the multivalley character
 of their energy landscapes. Here we use the damage spreading method (DS) to investigate
the cohesion of spin avalanches in the exponential networks and the scale-free
networks. On the contrary to the conventional methods, the results obtained from
DS suggest that the avalanche spectra are characterized by the same statistics
as the degree distribution in their home networks. Further, the obtained mean range $Z$
of an avalanche, i.e. the maximal distance reached by an avalanche from the damaged site, 
scales with the avalanche size $s$ as $Z/N^\beta =f(s/N^{\alpha})$, where $\alpha=0.5$ 
and $\beta=0.33$. These values are true for both kinds of networks for the number $M$ of 
nodes to which new nodes are attached between 4 and 10; a check for $M=25$ confirms 
these values as well.
\end{abstract}

\noindent


\noindent



\section{Introduction}


In recent years, there is a growing interest in research on complexity science. Common features
of physical, technical, biological, social and computational systems are qualified as constituting
 what we call complex systems. A final definition of a complex system would be premature.
What is repeated in different formulations is 'a system
built up from many interacting components'. Further, properties of a "complex system" are 'not fully
 explained by an understanding of its component parts' \cite{sci}. This notion allows to expect a
formulation of phenomenological (as opposed to 'ab initio') laws which could be valid at intermediate
levels of integration of system structure. The component parts can be reduced to simple mathematical
 objects, still the observed or derived laws can reveal new and unexpected aspects of a 'complex' system.\\

Archetypal examples of complex systems are growing networks \cite{doro}. They consist of nodes, and some
of these nodes are linked together. This generalization of the Cartesian lattices became an object
 of research of computational scientists in 1998, when Watts and Strogatz published their
seed paper \cite{ws}. Since then, the list of relevant references grown to thousands; for some recent
monographs we refer to \cite{b1,b2,b3,b4,b5,b6}. A specific branch of complexity emerges around networks 
decorated with Ising spins  $S_i=\pm 1$ at each node. There, the
interaction between linked spins can favorize the same (ferromagnet) or different (antiferromagnet, AF)
signs. This generalization is known \cite{doro} to enrich the list of problems and 
 possible applications of complex networks.\\

\begin{figure}[ht] 
\centering
{\centering \resizebox*{12cm}{9cm}{\rotatebox{00}{\includegraphics{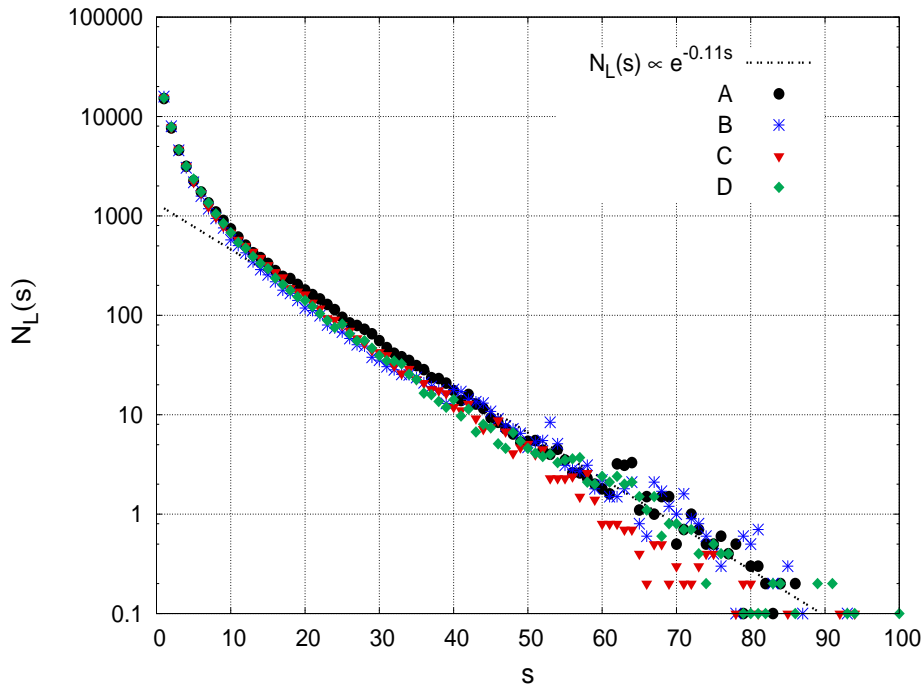}}}} 
\caption{
The avalanche spectrum $N_L(s)$ for the exponential networks, $M=5$.
}
\label{fig-1}
\end{figure}

\begin{figure}[ht] 
\centering
{\centering \resizebox*{12cm}{9cm}{\rotatebox{00}{\includegraphics{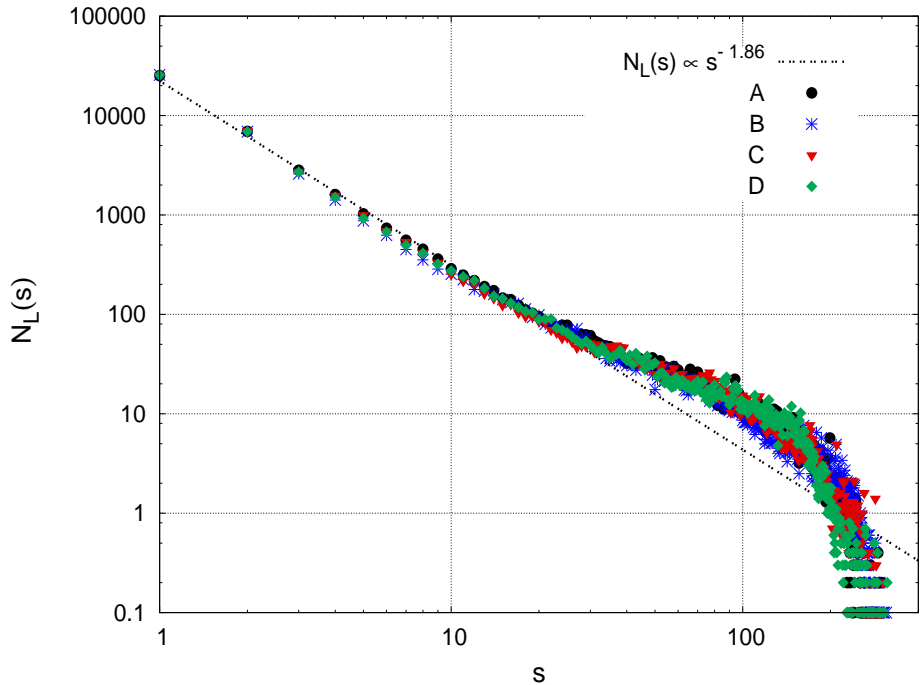}}}} 
\caption{
The avalanche spectrum $N_L(s)$ for the scale-free networks, $M=5$.
}
\label{fig-2}
\end{figure}

\begin{figure}[ht] 
\centering
{\centering \resizebox*{12cm}{9cm}{\rotatebox{00}{\includegraphics{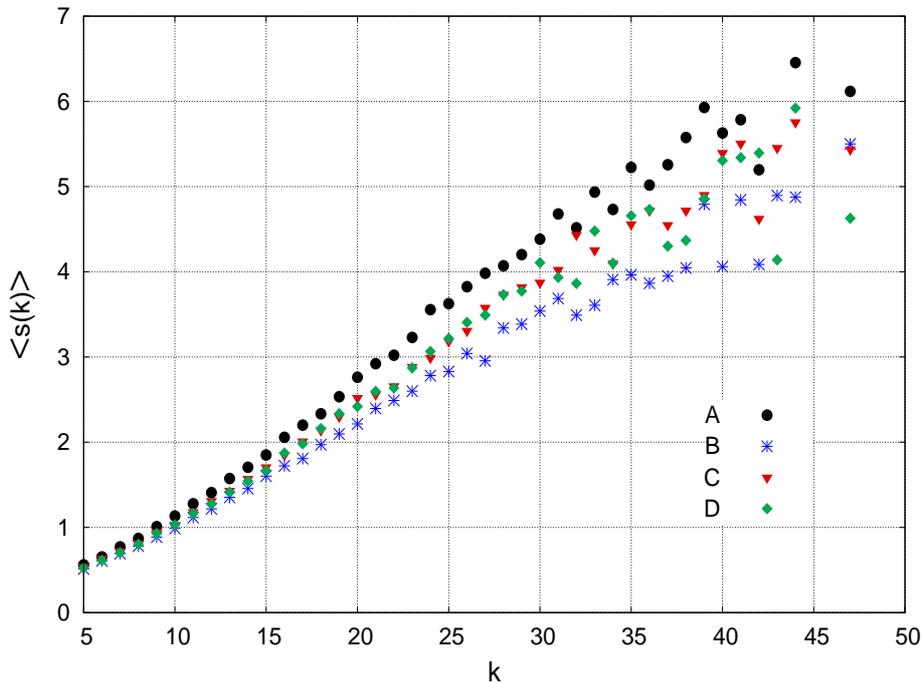}}}} 
\caption{
The average size $s$ of avalanche against the degree of the site where the avalanche was born, for the exponential networks, $M=5$.
}
\label{fig-3}
\end{figure}

\begin{figure}[ht] 
\centering
{\centering \resizebox*{12cm}{9cm}{\rotatebox{00}{\includegraphics{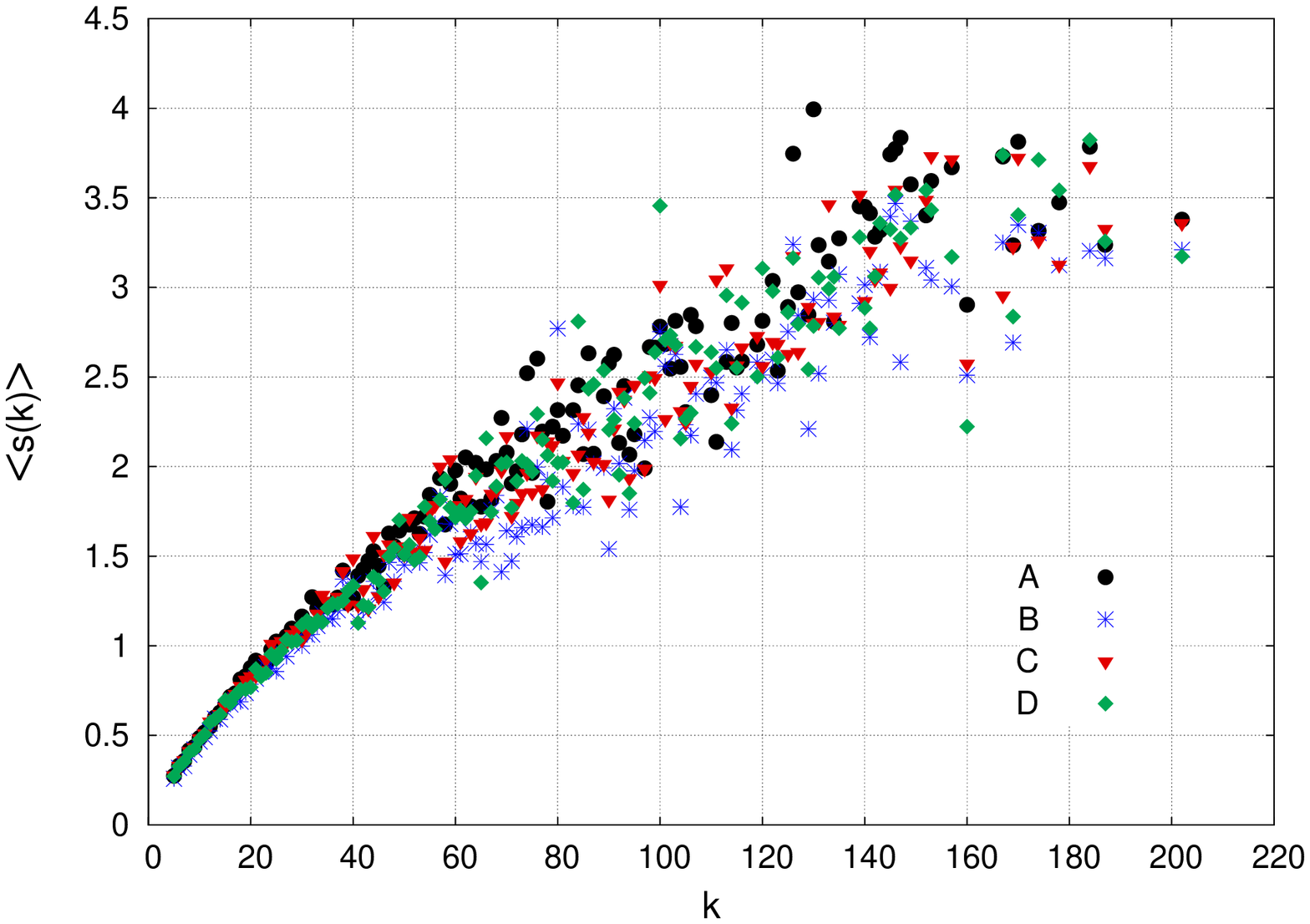}}}} 
\caption{
The average size $s$ of avalanche against the degree of the site where the avalanche was born, for the scale-free networks, $M=5$.
}
\label{fig-4}
\end{figure}
Here we are interested in statistics of avalanches in growing networks with AF interaction. The case of AF is
more complex, because this interaction leads to the effect of  frustration of spins in disordered 
 structures \cite{ksta}. Frustration means in particular that once three nodes are
connected to each other, spins at these nodes cannot have different signs; at least two spins must be
of the same signs, what raises the energy. Further, there are many states with the minimal
 (ground state) energy, while for a ferromagnet there are only two ground states. In frustrated networks, the
spectrum of ground states can be very rich and complex, and their number increases exponentially with the
system size \cite{mjk}. Counting and classification of these states is then computationally unfeasible.
 It seems that an information on the probability distribution of the number of states which lead to the same local minimum
can be related with the spectrum of connected avalanches. \\

The numerical experiment is conducted as follows.
 We vary the external field what leads to flipping of groups of spins. Two runs are performed with using the
same set of random numbers.  In one run we keep one spin blocked in the direction opposite to its direction 
in equilibrium before the field is changed. The number of spins which behave
 in a different way is the size $s$ of a connected avalanche. The method is known as the damage spreading technique
\cite{ds1,ds2}. In previous numerical experiments \cite{my1,my2} the size of an avalanches was calculated just
 as the number of flipped spins. That method did not differ between connected and disconnected avalanches.\\

The goal of this paper is twofold. First, the avalanche spectra in the growing networks are characterized by 
 the same statistics, as the degree distributions in these networks. This means, that the probability 
distribution of avalanche size $s$ in the exponential networks is exponential, and the one in the scale-free 
networks is the power function. Our second finding is that the obtained mean range $Z$ of avalanches, 
i.e. the maximal distance reached by an avalanche from the damaged site, scales with the avalanche size $s$ 
as $Z/N^\beta = f(s/N^{\alpha})$,  where $N$ is the number of nodes in the network. The exponents $\alpha$ 
and $\beta$ increase from zero for $M=1$ (trees) and they are approximately constant above some value of $M$.
Here $M$ is the growing parameter, i.e. the number of nodes to which new nodes are attached.
Above $M=4$, the values of $\alpha$ and $\beta$ are almost the same for the exponential and the scale-free networks. \\

In next section, the calculations and results are described in details. Section 3 is devoted to discussion.

\section{Calculations and results}

The scheme of our calculation is basically the same as in \cite{my1,my2}; the difference is that here we apply
the damage spreading technique. Avalanches are measured in the numerical experiment with the hysteresis loop.
Energy of spin $S_i$ is calculated as 

\begin{equation}
E_i=S_i\big[\sum_{j(i)}S_j-h]
\end{equation}
 where $h$ is the magnetic field energy. The energy units are $|J|$, where $J<0$ is the antiferromagnetic exchange
integral. The field is changed from $h_m$ to $-h_m$ by 1, where $h_m=k_{max}+0.5$, and $k_{max}$ is the maximal 
 degree in a given network. Each field change is performed $N$ times, where $N$ is the size of the network; in this 
way each spin is blocked once for the DS technique. Simulations are performed for four different orders of the sequence 
of nodes: (A) from the oldest to the newest nodes, (B) from the newest to the oldest nodes, (C) with random permutations, and (D) with one selected
sequence. The obtained plots indicate that these variants do not differ qualitatively. \\
 
The avalanche spectra, i.e. the number $N_L$ of avalanches of size $s$ for the exponential and the scale-free growing 
networks are shown in Fig. 1 and 2, respectively. These calculations are performed for the network size $N=2\times 10^3$ nodes, 
and the spectra are averaged over $K=10$ networks. The obtained relations for the exponential networks can be parametrized
as $N_L(s)\propto \exp(-\phi s)$, where $\phi=0.45$ for $M=1$, $\phi=0.21$ for $M=2$, and $\phi=0.11$ for $M=5$.
For the scale-free networks, the relation $N_L(s)\propto s^{-\gamma}$ applies. There, $\gamma=2.71$ for $M=1$, $\gamma=2.14$ 
for $M=2$, and $\gamma=1.86$ for $M=5$. \\

These results indicate that the avalanche spectrum is determined by degree distribution of the network. We checked how
the avalanche size depends of the degree of a node where the damage was started. The obtained plots are shown in Figs. 3 and 4,
for the exponential and the scale-free networks, respectively.\\

We investigate also the range $Z$ of avalanches, i.e. the maximal distance reached by an avalanche from the damaged site,
against the avalanche size $s$. Due to computational limitations, these calculations could be performed for the network size not larger
than $N=300$. However, here we got the finite-size scaling relations 

\begin{equation}
Z/N^{\beta}=f(s/N^{\alpha})
\end{equation}
with the shapes of the functions  $f(s)$ shown in Figs. 5-8 for $M=1$ and $M=5$, the exponential and the scale-free
networks. For $M=1$ results are averaged over $K=10^3$ networks ($N=300$) and over $K=10^4$ (for others $N$). For $M=5$ number of networks are $K=10^3$ except $N=300$ where $K=10^2$.The results on the obtained exponents $\alpha$ and $\beta$ against the growing parameter $M$ are shown in Fig. 9. 
These results indicate that $M=1$ is a special case, where the scaling relation does not depend on the network size $N$. 
On the contrary, for $M$ between 4 and 10 the plots do not depend qualitatively on $M$, and the values of the exponents 
$\alpha$ and $\beta$ seem to reach their asymptotic values. This is confirmed by the calculation for $M=25$, where we get
$\alpha=0.52(0.51)$ and $\beta=0.34(0.33)$ for the scale-free (exponential) networks. Most important result is that 
the exponents $\alpha$ and $\beta$ are the same for the exponential networks and the scale-free networks.

\begin{figure}[ht] 
\centering
{\centering \resizebox*{12cm}{9cm}{\rotatebox{00}{\includegraphics{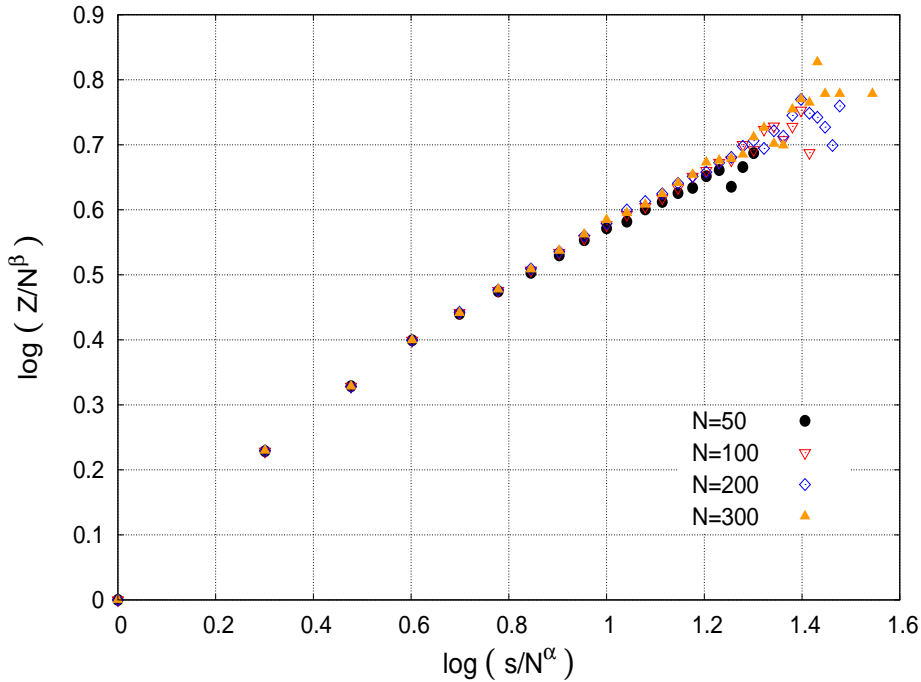}}}} 
\caption{
The scaling relation between the range $Z$ of avalanches and their size $s$ for the exponential trees ($M=1$).
}
\label{fig-5}
\end{figure}

\begin{figure}[ht] 
\centering
{\centering \resizebox*{12cm}{9cm}{\rotatebox{00}{\includegraphics{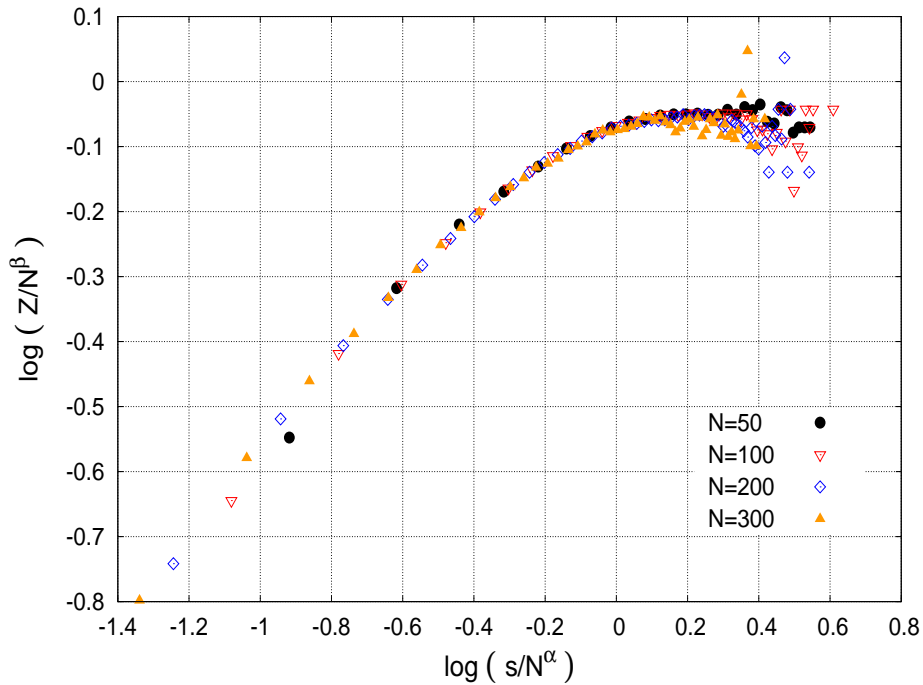}}}} 
\caption{
The scaling relation between the range $Z$ of avalanches and their size $s$ for the exponential networks, for $M=5$.
}
\label{fig-6}
\end{figure}

\begin{figure}[ht] 
\centering
{\centering \resizebox*{12cm}{9cm}{\rotatebox{00}{\includegraphics{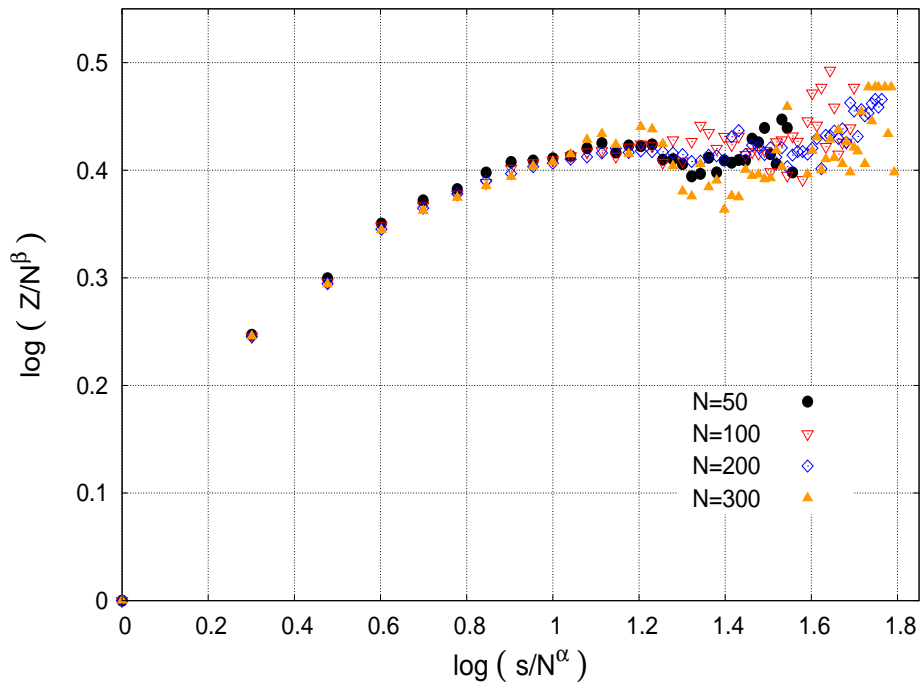}}}} 
\caption{
The scaling relation between the range $Z$ of avalanches and their size $s$ for the scale-free trees ($M=1$).
}
\label{fig-7}
\end{figure}

\begin{figure}[ht] 
\centering
{\centering \resizebox*{12cm}{9cm}{\rotatebox{00}{\includegraphics{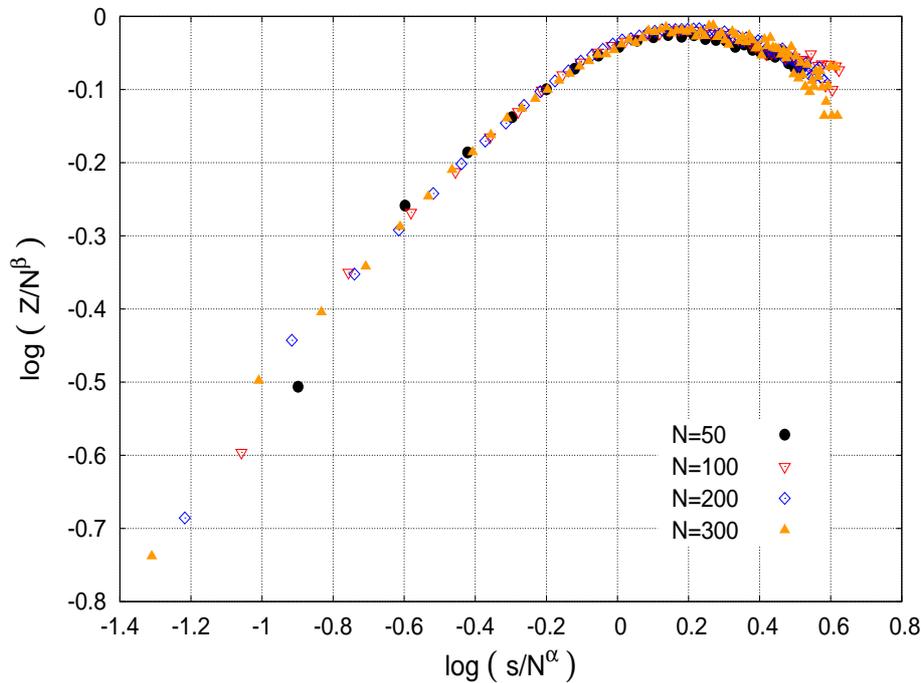}}}} 
\caption{
The scaling relation between the range $Z$ of avalanches and their size $s$ for the scale-free networks, for $M=5$.
}
\label{fig-8}
\end{figure}

\begin{figure}[ht] 
\centering
{\centering \resizebox*{12cm}{9cm}{\rotatebox{00}{\includegraphics{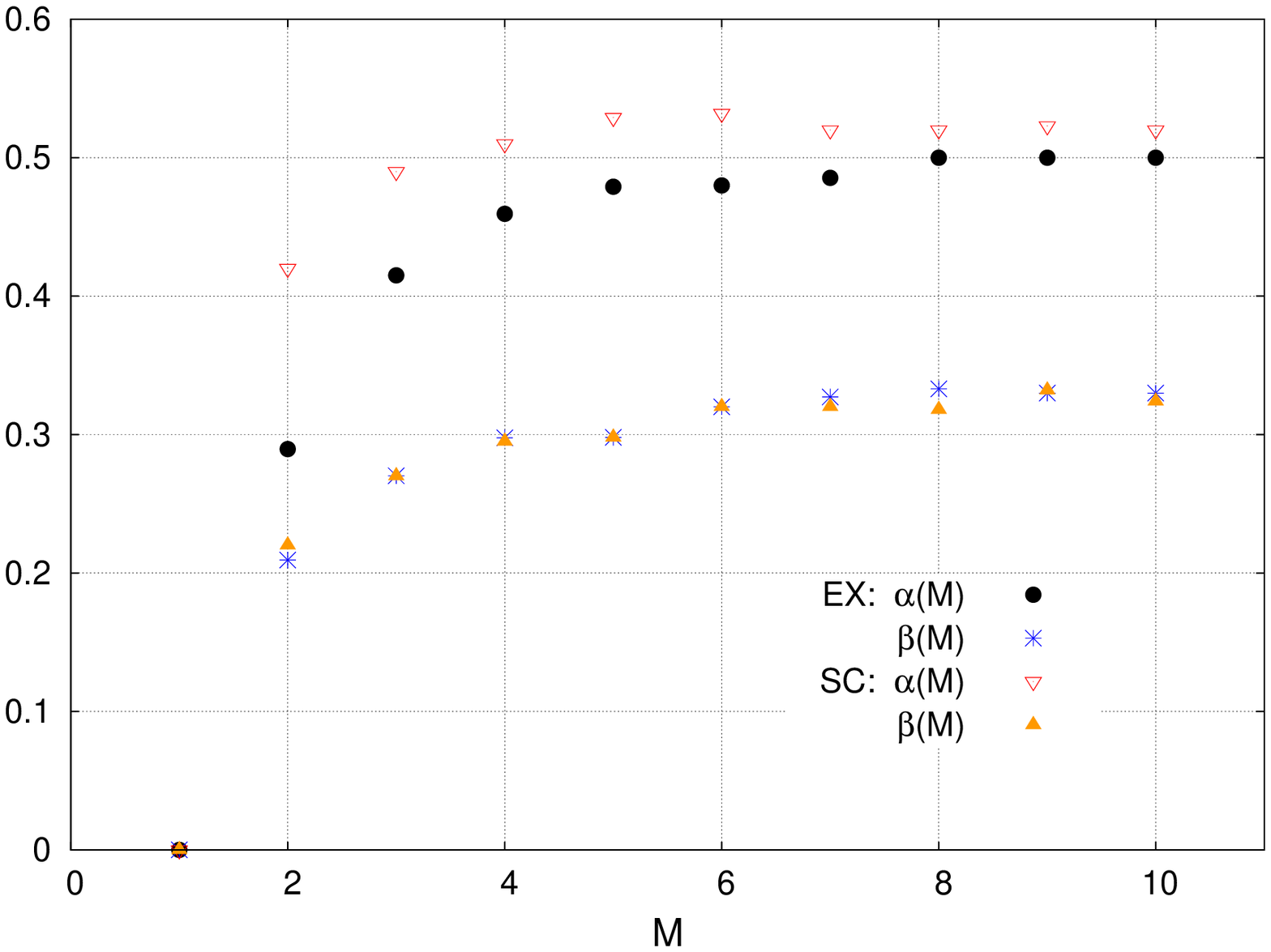}}}} 
\caption{
The exponents $\alpha$ and $\beta$ against the growing parameter $M$ for the exponential (EX) and the scale-free (SC) networks.
}
\label{fig-9}
\end{figure}

\section{Discussion}

Our numerical results indicate that the spectrum of avalanches is described by the same function as the degree distribution
in the growing network. A most simple explanation of this result could be that an avalanche contains all nodes in the direct 
neighborhood of a damaged node; the range of such avalanches is just one. We see that this is not the case; in general, the 
range of avalanche excesses 1. Still, the observed coincidence indicates that there is a monotoneous relation between the size 
of avalanche and the degree of the node at the avalanche origin. We checked numerically that indeed such a relation does appear. 
In other words, an avalanche born in a damaged site increases with the degree of this site. However, the obtained exponent $\gamma$ 
clearly decreases with the growing parameter $M$. This is in contradiction to the node degree distribution in the scale-free networks, 
where the appropriate exponent does not vary with $M$. Perhaps the finite size effect is enhanced here by the deviation of the relation
$<s(k)>$ from linearity, as observed in Fig. 4. Note that this argument is not related to the exponential networks, where the deviation 
from linearity (Fig. 3) is not observed, but the exponent $\phi$ is known to decrease with $M$ \cite{my2}.\\

The second result independent on the network topology is the scaling relation between the range and size 
of the avalanches, i.e.  $Z/N^{\beta}= f(s/N^{\alpha})$. For trees ($M=1$), the exponents $\alpha$ and $\beta$
vanish, what reflect the fact that in linear chains the range and the size of avalanches is the same. For $M=4$ and larger, 
$\alpha$ is close to 1/2 and $\beta$ is close to 1/3 with the numerical accuracy. This is true for the exponential and 
the scale-free networks. The plateau of the function $f(s)$, shown in Fig. 3, reveals that the range of avalanches is limited: 
The scaling relation found here suggests that the origin of this limitation is the size of the network. We are not aware of any 
analytical calculation of the range of avalanches. The topological disorder of the network, combined with the frustration, can 
produce a kind of magnetic disorder, analogous to the Random Field Model \cite{rfm,cardy}; perhaps some results of this model
could be applied also to the antiferromagnetic growing networks.

\bigskip

\section*{Acknowledgements}Project operated within the Foundation for Polish Science MPD Programme co-financed by the EU European Regional Development Fund.


\begin{thebibliography}{99}
\bibitem{sci} R. Gallagher and T. Appenzeller, {\it Beyond reductionism}, Science {\bf 284} (1998) 79.
\bibitem{doro} S. N. Dorogovtsev and A. V. Goltsev, {\it Critical phenomena in complex networks}, Rev. Mod. Phys. {\bf 80} (2008) 1275.
 \bibitem{ws} D. J. Watts and S. H. Strogatz, {\it Collective dynamics of 'small-world' networks}, Nature {\bf 393} (1998) 440.
\bibitem{b1} A.-L. Barabasi, {\it Linked: How Everything Is Connected to Everything Else and What
 It Means for Business, Science, and Everyday Life}, Plume Books, New York 2003.
\bibitem{b2} S. N. Dorogovtsev and J. F. F. Mendes, {\it Evolution of Networks: From Biological Nets to the Internet and WWW}, Oxford UP, Oxford 2003.
 \bibitem{b3} {\it Handbook of Graphs and Networks: From the Genome to the Internet}, Eds. S. Bornholdt and H. G. Schuster, Wiley-VCH, Berlin 2003.
\bibitem{b4} R. Pastor-Satorras and A. Vespignani, {\it Evolution and Structure of the Internet: A Statistical Physics Approach}, Cambridge UP, Cambridge 2004.
 \bibitem{b5} R. Durrett, {\it Random Graph Dynamics}, Cambridge UP, Cambridge 2006.
\bibitem{b6} G. Caldarelli, {\it Scale-Free Networks}, Oxford UP, Oxford 2007.
\bibitem{ksta} D. Stauffer and K. Ku{\l}akowski, {\it Why everything gets slower?}, TASK Quarterly {\bf 7} (2003) 257.
 \bibitem{mjk} M. J. Krawczyk, K. Malarz, B. Kawecka-Magiera, A. Z. Maksymowicz and K. Ku{\l}akowski, {\it Spin-glass properties of an Ising antiferromagnet on the Archimedean $(3,12^2$) lattice}, Phys. Rev. B {\bf 72} (2005) 24445.
 \bibitem{ds1} A. Coniglio, L. de Arcangelis, H. J. Herrmann and N. Jan, {\it Exact relations between damage spreading and thermodynamical properties}, Europhys. Lett. {\bf 8} (1989) 315.
\bibitem{ds2} M. L. Rubio Puzzo, E. V. Albano, {\it The damage spreading method in Monte Carlo simulations: a brief overview and applications to confined magnetic materials}, Commun. Comput. Phys. {\bf 4} (2008) 207.
 \bibitem{my1} B.Tadi\'c, K.Malarz and K.Ku{\l}akowski, {\it Magnetization reversal in spin patterns with complex geometry}, Phys. Rev. Lett. {\bf 94} (2005) 137204.
\bibitem{my2} K. Malarz, W. Antosiewicz, J. Karpi\'nska, K. Ku{\l}akowski and B. Tadi\'c, {\it Avalanches in complex spin networks}, Physica A {\bf 373} (2007) 785.
\bibitem{rfm} A. P. Young  (Ed.), {\it Spin Glasses and Random Fields}, World Scientific, Singapore 1997.
\bibitem{cardy} J. L. Cardy, {\it Random-field effects in site-disordered Ising antiferromagnets}, Phys. Rev. B {\bf 29} (1984) 505-507.
\end{thebibliography}
\end{document}